\documentclass[%
 reprint,
 amsmath,amssymb,
 aps,
]{revtex4-2}

\usepackage{graphicx}
\usepackage{dcolumn}
\usepackage{bm}
\usepackage{ascii}

\begin{document}

\preprint{APS/123-QED}

\title{Measurement of coherent elastic neutrino-nucleus scattering\\  from reactor antineutrinos}

\author{J. Colaresi$^{1}$}
\author{J.I. Collar$^{2}$}
\email{collar@uchicago.edu}
\author{T.W. Hossbach$^{3}$}
\author{C.M. Lewis$^{2}$}
\author{K.M. Yocum$^{1}$}
\affiliation{%
$^{1}$Mirion Technologies Canberra, 800 Research Parkway, Meriden, CT, 06450, USA
}%
\affiliation{%
$^{2}$Enrico Fermi Institute,
University of Chicago, Chicago, Illinois 60637, USA
}%
\affiliation{%
$^{3}$Pacific Northwest National Laboratory, Richland, Washington 99354, USA
}%


\date{\today}

\begin{abstract}
The 96.4 day exposure of a 3 kg ultra-low noise germanium detector to the high flux of  antineutrinos from a  power nuclear reactor is described. A very strong preference ($p<1.2\times10^{-3}$) for the presence of a coherent elastic neutrino-nucleus scattering (CE$\nu$NS) component in the data is found, when compared to a background-only model. No such effect is visible in 25 days of operation during reactor outages. The best-fit CE$\nu$NS signal is in good agreement with expectations based on a recent characterization of germanium response to sub-keV nuclear recoils. Deviations of order 60\% from the Standard Model CE$\nu$NS prediction can be excluded using present data. Standing uncertainties  in models of germanium quenching factor, neutrino energy spectrum, and background are examined.
\end{abstract}

\maketitle



We have recently reported \cite{dresden1} on the deployment of a p-type point contact (PPC) germanium detector \cite{ppc} in close proximity to the core of the Dresden-II boiling water reactor (BWR). The device, dubbed NCC-1701, combines presently unique characteristics of large mass (2.924 kg) and low energy threshold (0.2 keV$_{ee}$) within a compact shield, enabling a search for subtle sub-keV signals expected from  coherent elastic neutrino-nucleus scattering (CE$\nu$NS) \cite{freedman,science}. Details on detector, shielding, FPGA-based data-acquisition system (DAQ), background characterization, calibrations, and data treatment are provided in \cite{dresden1}. The main objective of this installation was to study the practicality of reactor monitoring using a small-footprint PPC assembly in the aggressive environment (radiation, temperature, EMI/RFI, acoustic noise, vibration) a few meters from a commercial  reactor core \cite{dresden1}. The neutrino detector miniaturization afforded by a large CE$\nu$NS cross-section permits to envision such technological applications, in a near future \cite{huber}.  Reactor sources can significantly expand the potential of CE$\nu$NS to probe physics beyond the Standard Model (SM) \cite{deviations}.

The new dataset highlighted in this Letter spans the period between 1/22/2021 and 5/8/2021, during which the reactor was operated at its full nominal power of 2.96 GW$_{th}$ (Rx-ON). Interruptions on days  515-516, 535-537, and 546-552 (referenced to  detector installation on 10/19/2019) were due to data storage overflows during a time of limited access to the site, resulting in a 96.4 day effective exposure. The start of this new run followed the installation of a spot cooler \cite{trip} able to  reduce the temperature inside the detector shield while external temperatures approached \mbox{35 C}. This  led to a significant decrease in detector cryocooler power and spurious DAQ triggers produced by its mechanical vibrations when operated in extreme conditions. An extended technical drop in reactor power determined the end of this run. 

\begin{figure}[!htbp]
\includegraphics[width=.87 \linewidth]{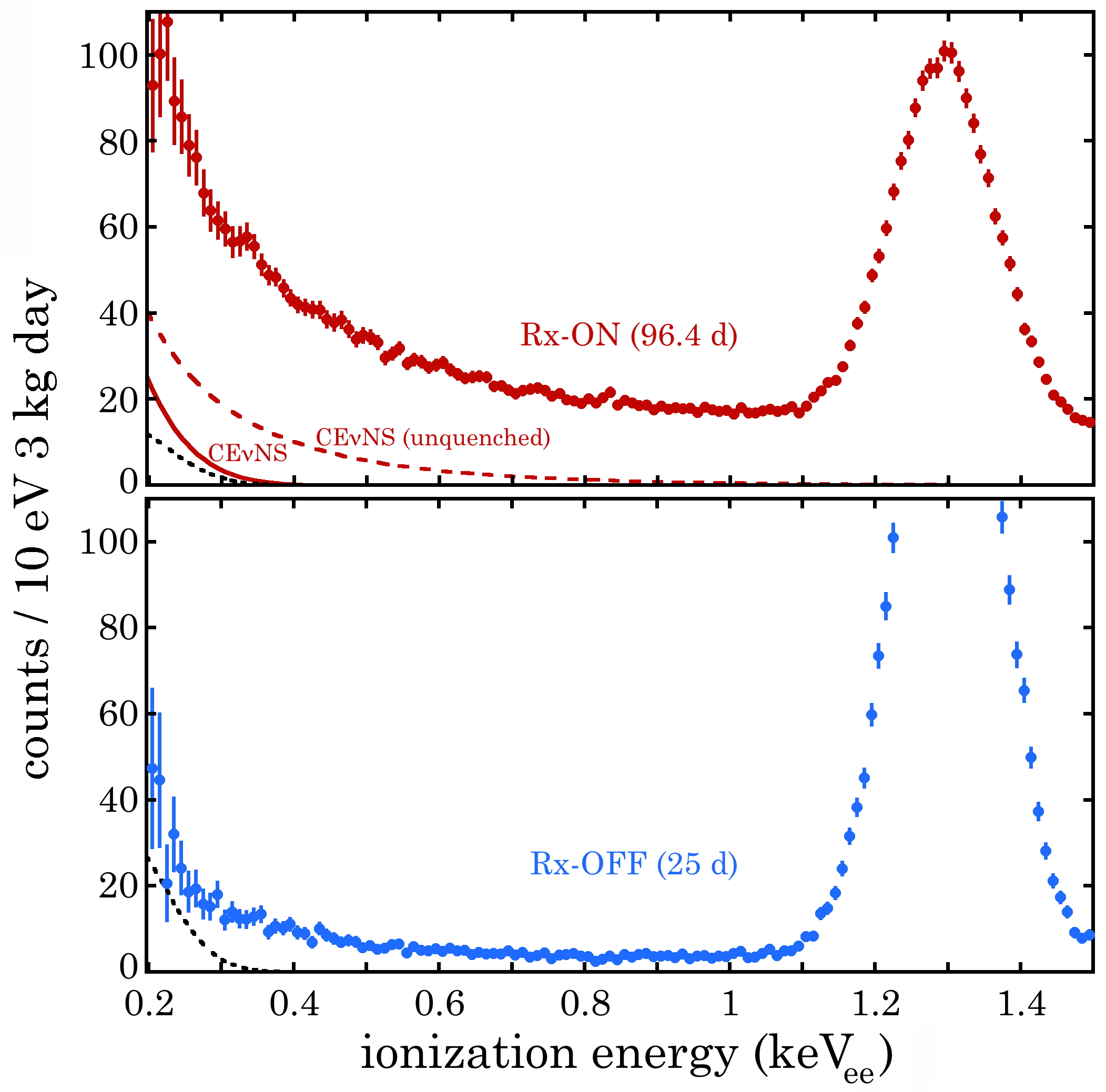}
\caption{\label{fig:epsart1} Energy spectra of  PPC bulk events during Rx-ON and Rx-OFF periods.  The  CE$\nu$NS expectation (red line) uses the MHVE antineutrino spectrum and Fe-filter quenching factor (see text). A dashed red line illustrates the impact of  quenching on CE$\nu$NS. Black dotted lines signal the   $^{71}$Ge M-shell EC contribution, derived from L-shell EC at 1.29 keV$_{ee}$. This process is noticeable in Rx-OFF data, taken prior to the addition of neutron moderator (i.e., following intense $^{71}$Ge activation).}
\end{figure}

An additional 2.5 cm-thick  layer of borated polyethylene was placed on the bottom side of the shield assembly on 6/13/2020. That surface featured only a minimal thickness of hydrogenated material, facilitating neutron ingress. As predicted by simulations using dedicated environmental background measurements as input \cite{dresden1}, this straightforward upgrade resulted in a further drop by a factor of two in the low-energy spectrum, presently dominated by the elastic scattering of epithermal neutrons \cite{dresden1}. The  peak at 1.297 keV$_{ee}$ from L-shell electron capture (EC) in $^{71}$Ge following neutron capture in $^{70}$Ge was also reduced by a factor of four with respect to its previous activity \cite{dresden1}, once a new secular equilibrium between $^{71}$Ge production and decay (T$_{1/2}$=11.4 d) was reached. The 20 day Rx-ON run reported on in \cite{dresden1} immediately followed a first installation of neutron moderator, i.e., was pre-equilibrium and hence subject to an elevated $^{71}$Ge decay rate from preceding activation. 

New information obtained from the reactor operator  allowed to  establish a precise distance between PPC crystal and the center point of the BWR core. This had been estimated in \cite{dresden1} at 8 m based on initially available data. The improved figure is a center-to-center distance of 10.39 m, with active fuel elements located 7.48 m to 13.31 m from the PPC. The axial and radial power profiles of the fuel assembly, specific to this period of exposure, were used to simulate the effect of an extended antineutrino source (3.66 m tall, 4.57 m diameter) in such close proximity to a comparatively point-like target. This produces a negligible 0.8\% reduction in the antineutrino flux expected from a point-equivalent source at 10.39 m. Our best-effort estimate of this flux is then 4.8$\times10^{13}~\bar{\nu_{e}}/$cm$^{2}$s with a $\sim$2\% uncertainty based on the dispersion seen in other assessments \cite{wong,connie,gemma,conus2}. Time-dependent changes of O(0.1)\% to this flux during the fuel cycle \cite{wong} are neglected here.

\begin{figure}[!htbp]
\includegraphics[width=.87 \linewidth]{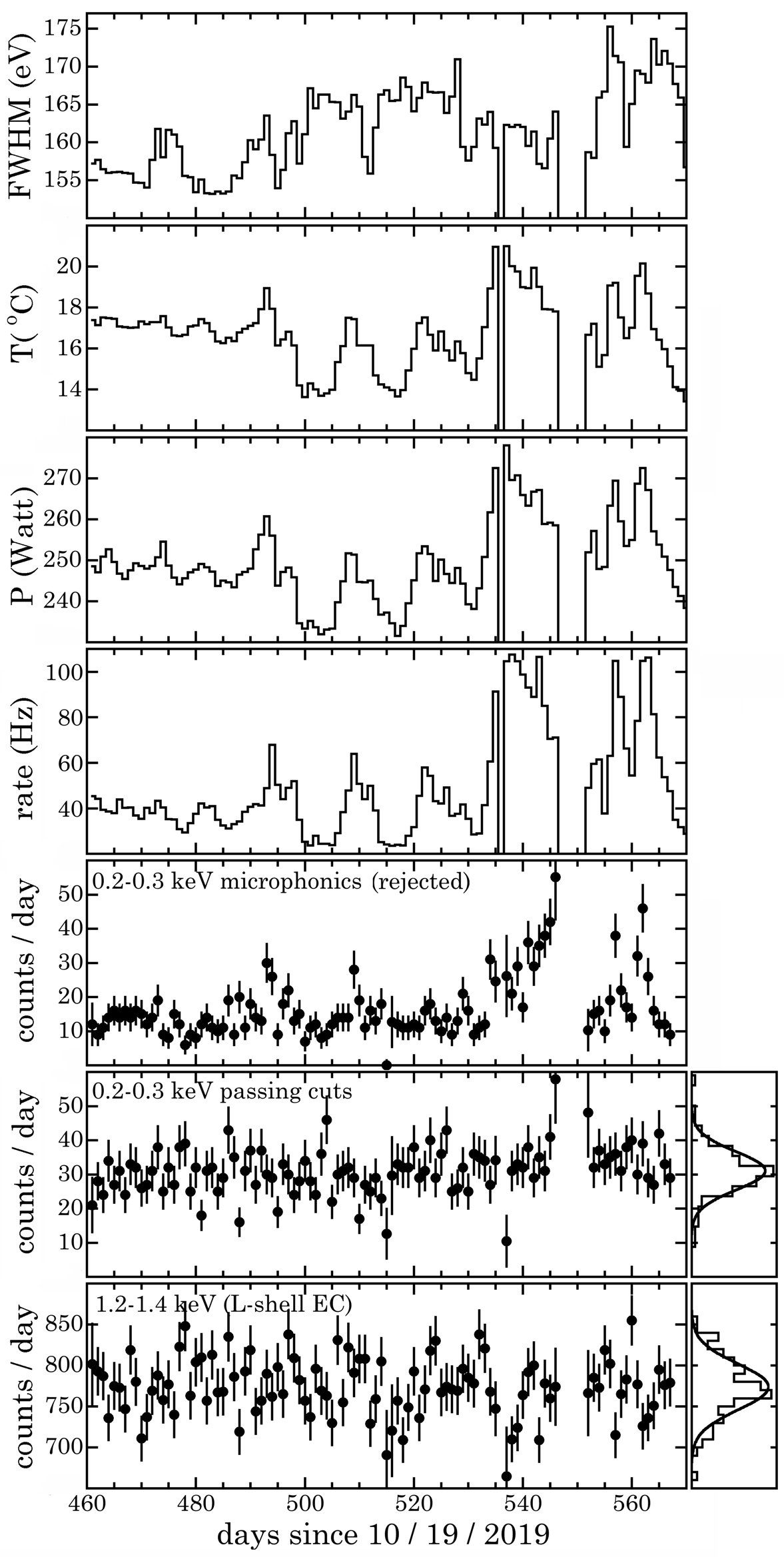}
\caption{\label{fig:epsart2}  From top to bottom (panels 1-4 are daily averages, 5-7 daily rates): 1) PPC electronic noise, measured using pre-trigger preamplifier traces, 2) temperature inside the shield, 3) cryocooler power, 4) DAQ trigger rate, 5) events rejected by a last-applied quality cut against microphonics, 6) events passing all cuts, 7) events under the L-shell EC peak. Error bars are statistical and therefore larger for partial data-acquisition days (see text). Side panels show data dispersion (histogram) and the  Gaussian expected from their mean. }
\end{figure}

A third {\it in situ} characterization of detector response to low-energy signals was performed on 1/22/2021, using a programmable electronic pulser to mimic the fast rise-time of preamplifier signals from interactions in the bulk of the PPC crystal \cite{dresden1,surface}. This confirmed the stability of signal acceptance (SA) over extended periods of time \cite{release}. The uncertainty associated to SA from the cumulative of these pulser calibrations varies from 8.6\% to 1.3\% across the sub-keV region of interest (ROI) for a CE$\nu$NS search \cite{release}. This is combined with the statistical uncertainty of events passing data cuts, to generate  error bars in the reconstructed (i.e., corrected for SA) spectra of  \mbox{Fig.\ 1. }

The choice of stringent data cuts made in  \cite{dresden1} was maintained in the present analysis, with the exception of a small increase to an edge-finding condition ($\varepsilon>\varepsilon_{min}$, Fig.\ 2 in \cite{dresden1}). This conservative measure was preemptively adopted to ensure an absence of  correlation between the accepted event rate near detector threshold, where the CE$\nu$NS signal is expected to accumulate, and  time-varying systematics discussed next. For consistency, the data previously-acquired  during 25 days of reactor outages \cite{dresden1} were re-analyzed with this modification. Resulting changes to this Rx-OFF spectrum (Fig.\ 1) are minimal and dominated by a small drop in overall normalization, traceable to an incorrectly calculated dead-time fraction in \cite{dresden1}. Both spectra in Fig.\ 1 account for  dead-time  (16.5\%) from spurious veto coincidences and PPC preamplifier saturation.

Special attention was paid to testing for contamination (unrejected electronic or microphonic noise,  slow rise-time surface events \cite{surface}) in the spectral region next to the 0.2 keV$_{ee}$ analysis threshold, as this might lead to an excess able to mimic a CE$\nu$NS signal. Fig.\ 2 displays the temporal evolution during  Rx-ON  of environmental parameters able to cause such backgrounds, together with daily rates of low-energy (0.2-0.3 keV$_{ee}$) events rejected by quality cuts \cite{dresden1}, those passing all cuts, and for signals contributing to the L-shell EC peak. Four statistical estimators (\mbox{Blomqvist $\beta$}, Goodman-Kruskal $\gamma$, Kendal $\tau$, Spearman Rank) were used to explore monotonic -but not necessarily linear- correlations between these non-normal datasets. The implementation of these tests used here \cite{wolfram} generates a $p$-value statistic, with $p<0.05$ signaling that the hypothesis of independence between datasets is unlikely. As expected from considerations expressed above, temperature, cryocooler power, and trigger rate exhibit a strong correlation ($10^{-23}<p<10^{-14}$) evident by simple inspection of Fig.\ 2. The rate of low-energy events rejected by quality cuts clearly correlates to these environmental parameters ($4\cdot10^{-11}<p<10^{-3}$). Low-energy events passing all cuts are on the other hand independent of  environmental factors ($0.13<p<1$), including electronic noise, and as such exhibit the expected Gaussianity around their mean. They are also independent of  rejected events ($0.38<p<0.84$). The rate of L-shell EC events is independent of trigger rate ($0.41<p<0.45$), illustrating the stability of DAQ throughput during this run. The electronic noise intrinsic to the detector is independent from all other environmental factors ($0.09<p<0.33$), indicating a good stability of PPC leakage current in present temperature conditions \cite{dresden1}. This type of correlation analysis is of crucial importance prior to a CE$\nu$NS search using PPCs, in view of the  dominance of  microphonic noise in the ROI \cite{thesis}.   

A possible contamination with surface events in the spectra of Fig.\ 1 was quantified by studying the rise-time distributions of Rx-ON signals passing all cuts,  prior to the application of a stringent condition selecting fast bulk events \cite{dresden1}. Unconstrained, characteristic lognormal distributions \cite{dresden1,cogent,cdexrt} were used for fitting slow and fast components. Fits were performed as a function of energy, over 50 eV-wide regions. Following the rise-time cut, the contamination derived from these fits is a negligible 1.5\%, with no tendency to increase near threshold. 

\begin{figure}[!htbp]
\includegraphics[width=.87 \linewidth]{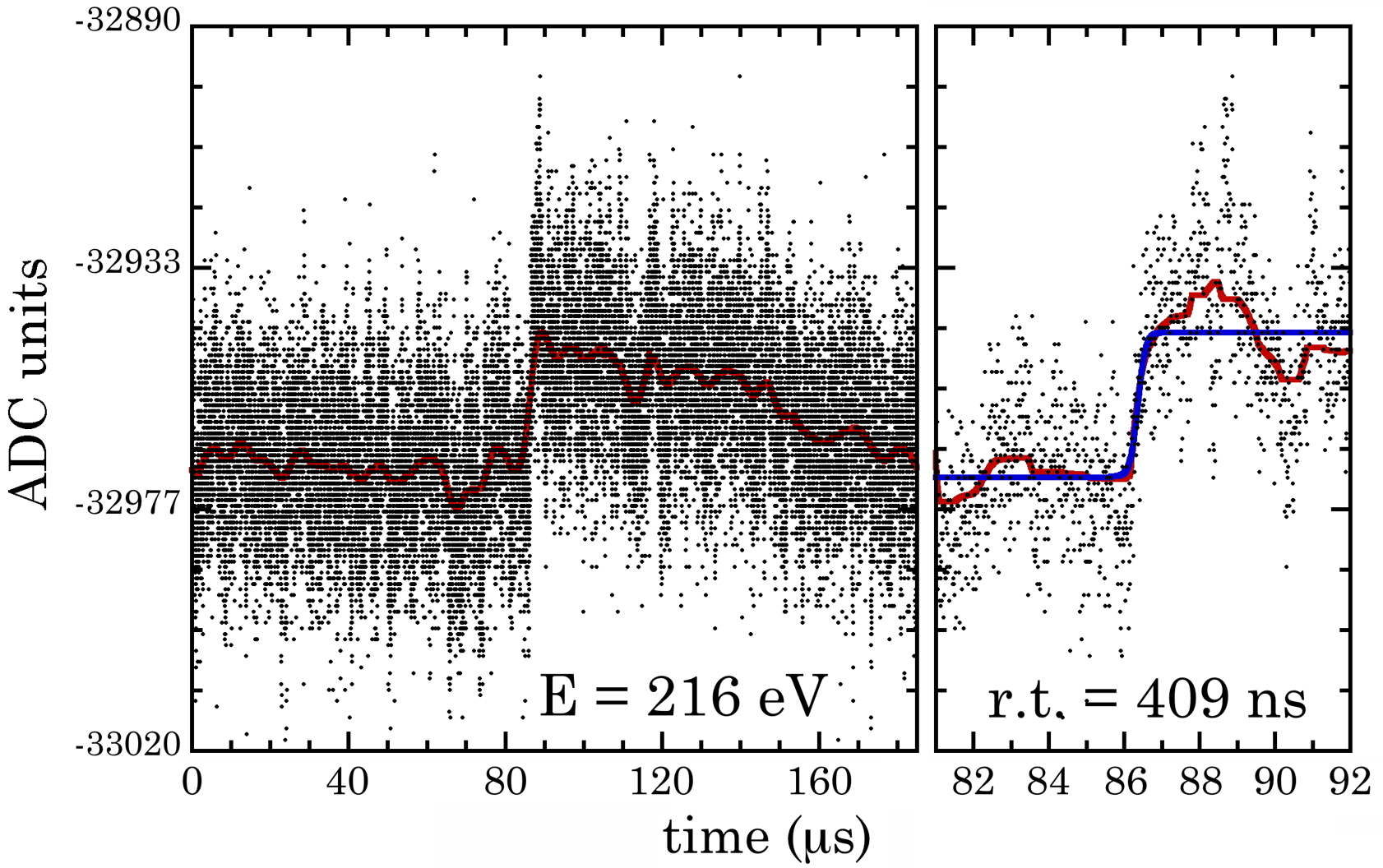}
\caption{\label{fig:epsart3}  Preamplifier trace for a typical radiation-induced low energy pulse passing all cuts, showing characteristic rise- and decay-times. The latter is intentionally elongated \cite{dresden1} via a 1 $\mu$F DC-blocking capacitor. Red lines show the wavelet-denoised trace. Wavelet parameters are separately optimized for edge-finding (left panel, \cite{dresden1}) and rise-time (r.t.) analysis (right panel). A blue line shows a tanh fit to the rising edge \cite{cdextanh}. A strict   condition (r.t.$<$ 660 ns, \cite{dresden1}) selects PPC bulk events with negligible surface event contamination (see text). }
\end{figure}

As a final safeguard,  traces from a large fraction (25\%) of events below 0.275 keV$_{ee}$ passing all cuts were randomly selected for visual inspection (Fig.\ 3). No candidates suspect of failing rejection of surface events  or any other anomalies were found. This last precaution, made possible by a custom  FPGA-based DAQ \cite{dresden1} is also important: commercial DAQ systems unable to digitize and store individual  preamplifier traces \cite{thesis,lastconus1,lastconus2} are not indicated for a CE$\nu$NS search,  as they do not allow to reject surface events or temporally-isolated microphonics, both prone to near-threshold accumulation.  

Fig.\ 1 displays the CE$\nu$NS signal predicted by the SM for the present Rx-ON exposure. Two antineutrino energy spectra were considered for this calculation. The first (``Kopeikin" in what follows), previously used in \cite{dresden1}, is described in \cite{kopeikin}. The second (``MHVE") adopts an approach  suggested in \cite{review}, where spectral information above a neutrino energy of 2 MeV is taken from \cite{huber1,mueller}, and \cite{vogel} is used below this energy. We find good agreement between our signal predictions and those by others \cite{huber,danny,dimitris}, but note small differences of order 20\% among these. Similarly for the spectral choices just described. The need for a standarized reactor source description has been recently emphasized \cite{wondram}. 

The CE$\nu$NS signal-to-background ratio achieved at threshold is $\sim$1/4, same as during the first observation of this process, prior to the subtraction of environmental backgrounds  feasible at a pulsed spallation neutrino source \cite{science,bjorn}.  For a steady-state reactor source this  is not a possibility. Instead, a spectral model able to describe sub-keV backgrounds must be adopted to investigate a CE$\nu$NS signal presence. Large, predictable spectral changes following sequential additions of very modest amounts of neutron moderator (a factor of three reduction in sub-keV rate in \cite{dresden1}, a factor of two here) point to the elastic scattering of epithermal neutrons as the presently dominant source. The  spectral shape model used to describe this component (a free exponential plus a free constant term \cite{dresden1}) is simple, as expected for a  ROI just $\sim$1 keV wide. It is also versatile, accommodating  simulated variations in epithermal spectral hardness, cross-section libraries, quenching factor -including deviations from the Lindhard model \cite{lindhard,qf}-,  veto threshold, and progressive addition of moderator \cite{dresden1,release}. 
 
The quenching factor (QF) describes the observed reduction in ionization yield produced by a nuclear recoil when compared to an electron recoil of same energy (Fig.\ 1, \cite{qf}). The CONUS experiment has recently imposed bounds on germanium QF models applicable to CE$\nu$NS \cite{lastconus1}.  Those still allowed generate a negligible CE$\nu$NS contribution to the NCC-1701 Rx-ON spectrum above $\sim$0.45 keV$_{ee}$. Applying the epithermal background model to the spectra of Fig.\ 1 above this energy generates an excellent fit to both Rx-ON and Rx-OFF data. However, extrapolation of the best fit  down to threshold points to a significant low-energy excess in both cases.  Electron capture from the M-shell in $^{71}$Ge  can contribute to this excess via the effect of detector energy resolution on its expected 0.158 keV$_{ee}$ deposition (Fig.\ 1, \cite{dresden1}). The ratio of M-shell to L-shell EC in germanium has been experimentally determined at 0.16$\pm$0.03 \cite{supercdms}, in good agreement with a theoretical expectation of 0.17 \cite{theory}. Using this ratio and extending the fitting region to encompass L-shell EC generates a robust prediction for the M-shell contribution (Fig.\ 1, \cite{release}).  An increasing SA uncertainty above 1.5 keV$_{ee}$ \cite{release} defines a fitting window extending from  threshold to this energy. A Gaussian L-shell EC peak with free amplitude, centroid and width, including a subsidiary contribution from L$_{2}$ capture \cite{mougeot1,mougeot2,release}, completes the background model.  

A Markov Chain Monte Carlo (MCMC) ensemble sampler \cite{mcmc1,mcmc2} was employed to fit the null hypothesis $\mathrm{H}_{0}$ (background model only) and several alternative hypotheses $\mathrm{H}_{1}$ (background plus CE$\nu$NS signal, with a dependence on choice of QF and neutrino spectrum) to Rx-ON and Rx-OFF spectra, treating both datasets identically. The preference for one hypothesis over the other is obtained via their Bayes factor ($\mathrm{B}_{10}$, ``ratio of evidences"), computed by evaluating their respective integrals over  parameter likelihood and assuming the same prior probability for both \cite{bayes}. This approach to hypothesis testing allows to compare non-nested models with the same number of degrees of freedom \cite{release}, as is done below, and to rank the premises used to define H1.

\begin{figure}[!htbp]
\includegraphics[width=.87 \linewidth]{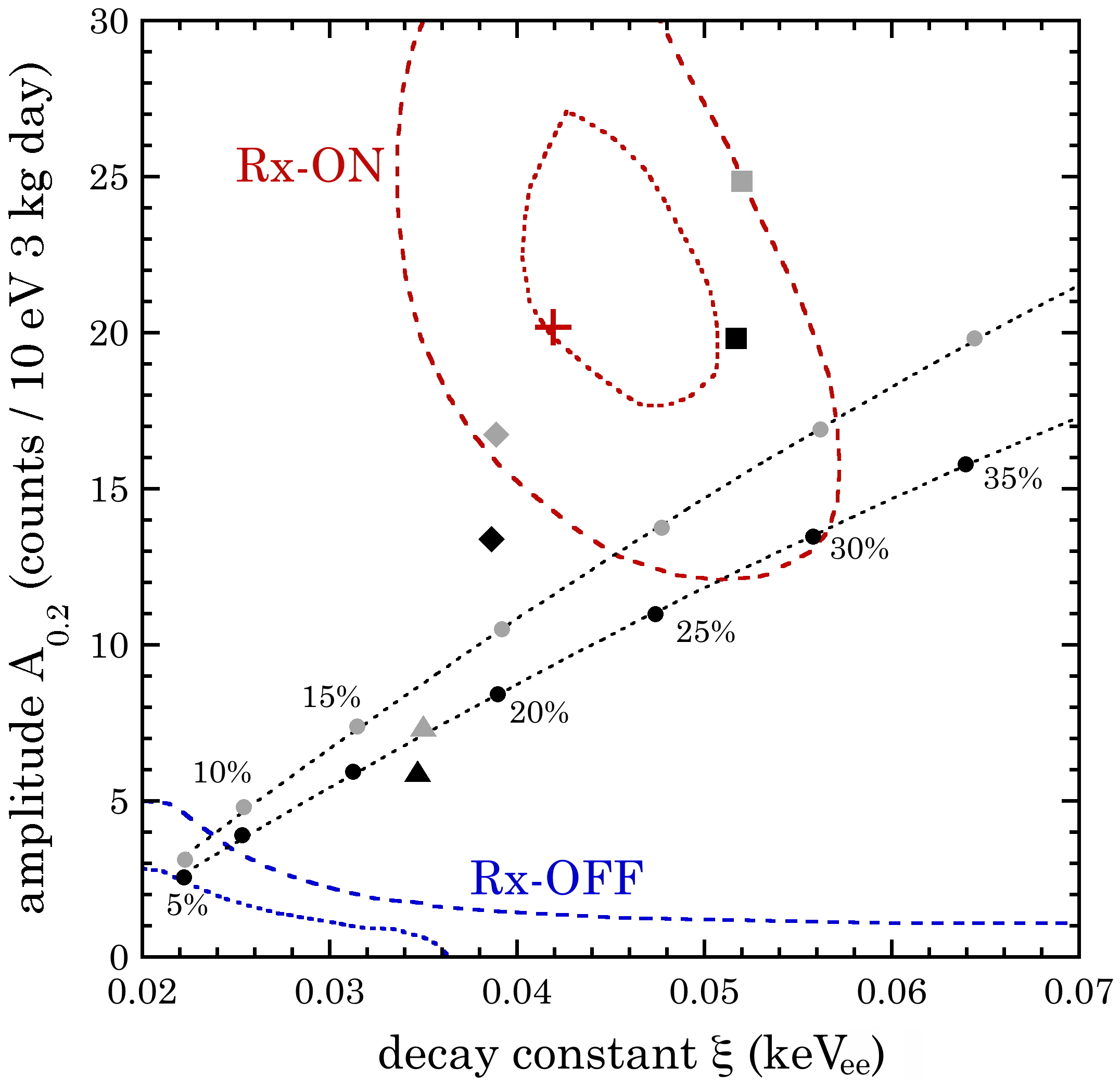}
\caption{\label{fig:epsart4} Favored values for CE$\nu$NS parameters $A_{0.2}$ and $\xi$ for Rx-ON (red) and Rx-OFF (blue) data. Dotted (dashed) lines indicate the 1-$\sigma$ (2-$\sigma$) contour extracted from MCMC corner plots \cite{release}. A cross marks the best-fit Rx-ON parameter values. For Rx-OFF this is $A_{0.2}=3.6^{+6.2}_{-2.7}$, with ill-defined $\xi$. Other symbols denote CE$\nu$NS predictions for combinations of neutrino spectra (gray for MHVE, black for Kopeikin) and QF (circles for indicated constant values, triangles for Lindhard with $\kappa$ = 0.157 \cite{release}, diamonds for YBe, and squares for Fef).}
\end{figure}

In a first step the CE$\nu$NS signal in $\mathrm{H}_{1}$ was approximated as an exponential  $A_{0.2}~ e^{-(E-0.2)/ \xi }$, where $A_{0.2}$ is its amplitude at threshold, $E$ is energy in keV$_{ee}$, and $\xi$ a decay constant. This parametrization is introduced solely to investigate which QF models are best supported by the data. The quality of this approximation is adequate, but varies  across QF models. Introducing a prior to account for CONUS constraints on the QF \cite{release}, the best fit values of $A_{0.2}$ and $\xi$ for the Rx-ON spectrum are in good agreement with expectations based on a recent QF characterization using sub-keV nuclear recoils (\cite{qf}, Fig.\ 4). The QF models favored are not in tension with present CONUS data \cite{release}. These are denoted by ``YBe" for a model based on photoneutron source measurements and ``Fef" for one derived using iron-filtered monochromatic neutrons \cite{qf,release}.   As expected, the best-fit  $A_{0.2}$ for Rx-OFF is compatible with zero (Fig.\ 4). Fig.\ 5 illustrates the good agreement of the background model with both Rx-ON and Rx-OFF data (reduced chi-square of 0.90 and 0.87, respectively), and the presence of a CE$\nu$NS-compatible excess for Rx-ON only. 

\begin{figure}[!htbp]
\includegraphics[width=.87 \linewidth]{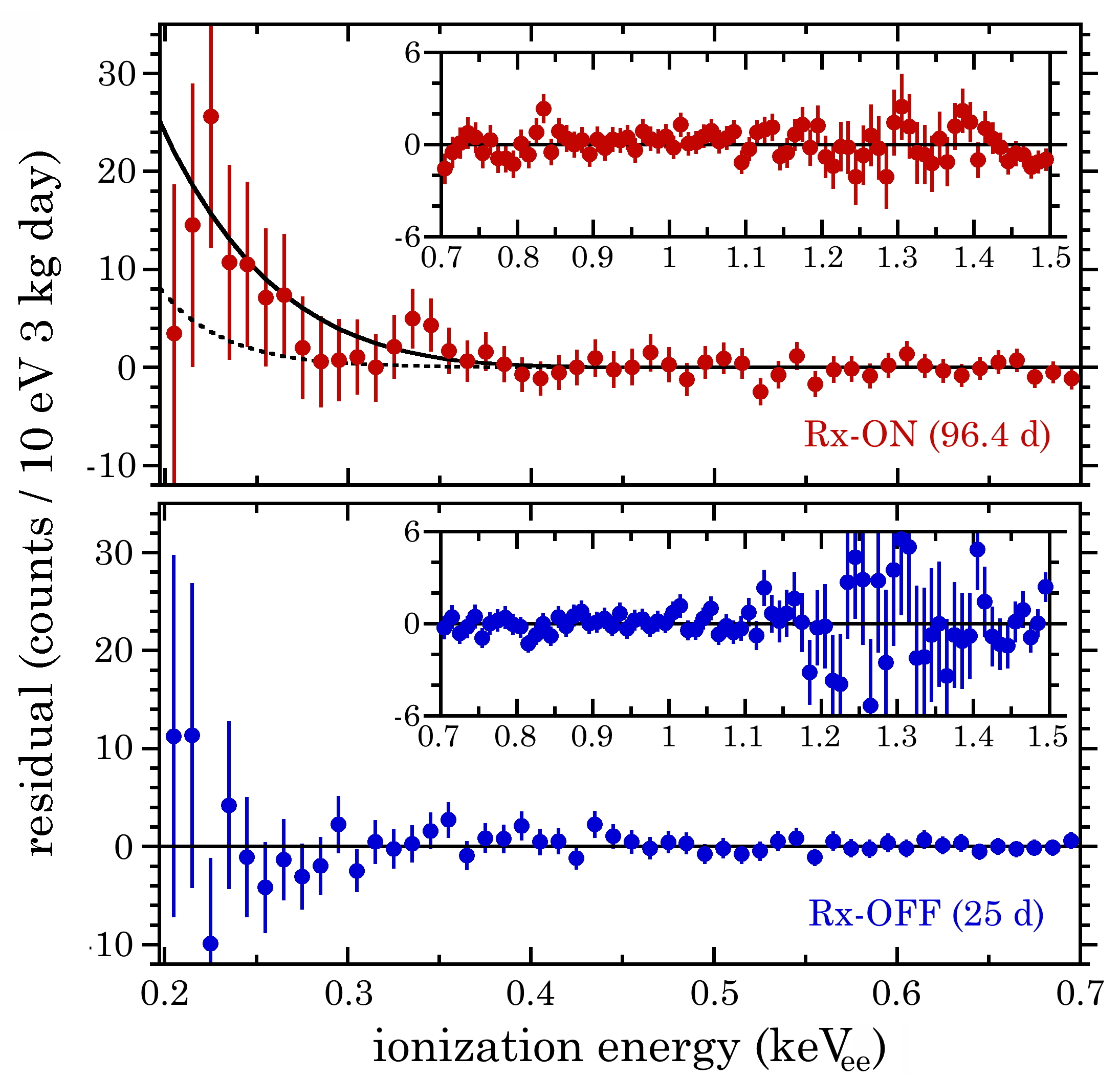}
\caption{\label{fig:epsart5}  Residual difference between the spectra of Fig.\ 1 and the best-fit background components (epithermal neutron, M-, L$_{1}$-, L$_{2}$-shell EC) in the alternative hypothesis $\mathrm{H}_{1}$ when using an exponential approximation for the CE$\nu$NS signal (see text). A solid (dotted) line shows the calculated CE$\nu$NS  signal prediction for MHVE-Fef (MHVE-Lindhard) in the SM. }
\end{figure}

The preference for $\mathrm{H}_{1}$ over $\mathrm{H}_{0}$ (i.e., rejection of $\mathrm{H}_{0}$ in favor of $\mathrm{H}_{1}$) was quantified for  six combinations of QF and neutrino spectrum (triangles, diamonds and squares in Fig.\ 4), without free parameters or approximations in the CE$\nu$NS component. For Rx-ON, Fef yields a $\mathrm{B}_{10}$ equal to 34.0 (MHVE) and 34.8 (Kopeikin). This corresponds to a ``very strong" preference for the presence of CE$\nu$NS according to the classic tabulation by Jeffreys \cite{jeffreys}. For the YBe QF this becomes $\mathrm{B}_{10}=$13.2 (MHVE) or 11.2 (Kopeikin), i.e., ``strong" evidence for  $\mathrm{H}_{1}$. For Lindhard the values are 4.0 and 3.1, respectively (``moderate" Bayesian evidence).  All $\mathrm{B}_{10}$ values for Rx-OFF support $\mathrm{H}_{0}$ instead ($\mathrm{B}_{10}\ll1$), ranging from ``moderately" (Lindhard) to ``extremely" (Fef).

As a final test, a free amplitude was allowed to multiply the SM differential rate calculated for the favored MHVE-Fef (Kopeikin-Fef) interpretations, during $\mathrm{H}_{1}$ fits. Its returned best-fit value is $0.97^{+0.31}_{-0.27}$ ($1.17^{+0.42}_{-0.42}$). Deviations from the SM introducing new physics involve CE$\nu$NS spectral distortions more complex than a mere change in signal rate \cite{deviations}. However, based on this simple assessment, the present dataset \cite{release} should allow to exclude deviations of order 60\% with $\sim$95\% confidence, as long as the new QF models in \cite{qf} are embraced. This best-fit amplitude is nevertheless $3.2^{+0.14}_{-0.15}$ for MHVE-Lindhard, as can be ascertained from Fig.\ 5. If Lindhard is accepted as an accurate description of the sub-keV germanium QF, this might  be interpreted as  evidence for new physics \cite{danny} or alternatively, for the incompleteness of the background model employed. Comparisons over the 0.45-1.5 keV$_{ee}$ spectral region of the present $\mathrm{H}_{0}$ against alternatives involving more complex  background spectral shapes (exponential plus linear, double exponential, \frenchspacing{etc.}) moderately but consistently favor the current $\mathrm{H}_{0}$ description under a Bayes factor test. Further work in understanding sub-keV quenching factors and their underlaying physical mechanisms is needed, for CE$\nu$NS to meet its full potential in the search for  physics beyond the SM \cite{csiqf,qf,danny}.

Constrained by the caveats that present uncertainties in quenching factor \cite{conusqf,ourcom}, background model   and  antineutrino spectrum introduce, this dataset provides a very strong preference for an interpretation that includes the Standard Model CE$\nu$NS signal, present during periods of reactor operation only. Experimentation with NCC-1701 at the Dresden-II BWR has provided an invaluable operating experience: with additional shielding upgrades that preserve compactness and the resolution of a technical issue  affecting  the  neutron veto \cite{dresden1}, Rx-ON backgrounds can be made comparable to Rx-OFF even in closest-possible proximity to a reactor core (simulations show that full veto performance alone would reduce the Rx-ON background by a further factor of two). Based on previous experience at sites profiting from a shallow overburden \cite{phil,ESS}, a planned relocation of NCC-1701 to a ``tendon gallery" surrounding a reactor containment dome should generate a signal-to-background ratio at threshold of \mbox{$\gtrsim$ 20}. This will facilitate a precision measurement of CE$\nu$NS and PPC utilization as a reactor-monitoring tool. An exciting future of applications in fundamental science and technology awaits for CE$\nu$NS detectors.

This work was supported by awards DARPA W911NF1810222 and NSF PHY-1812702. We are deeply grateful to Exelon Corporation for the generosity demonstrated in providing access to the Dresden-II reactor, as well as for their assistance and supervision in numerous instances. We are similarly indebted to Alex Kavner and Alan Robinson for their participation in  early stages of this project and to Luca Grandi, Xavier Mougeot and Dimitris Papoulias for helpful exchanges.

\bibliography{apssamp}

\providecommand{\noopsort}[1]{}\providecommand{\singleletter}[1]{#1}%
\begin{thebibliography}{55}%
\makeatletter
\providecommand \@ifxundefined [1]{%
 \@ifx{#1\undefined}
}%
\providecommand \@ifnum [1]{%
 \ifnum #1\expandafter \@firstoftwo
 \else \expandafter \@secondoftwo
 \fi
}%
\providecommand \@ifx [1]{%
 \ifx #1\expandafter \@firstoftwo
 \else \expandafter \@secondoftwo
 \fi
}%
\providecommand \natexlab [1]{#1}%
\providecommand \enquote  [1]{``#1''}%
\providecommand \bibnamefont  [1]{#1}%
\providecommand \bibfnamefont [1]{#1}%
\providecommand \citenamefont [1]{#1}%
\providecommand \href@noop [0]{\@secondoftwo}%
\providecommand \href [0]{\begingroup \@sanitize@url \@href}%
\providecommand \@href[1]{\@@startlink{#1}\@@href}%
\providecommand \@@href[1]{\endgroup#1\@@endlink}%
\providecommand \@sanitize@url [0]{\catcode `\\12\catcode `\$12\catcode
  `\&12\catcode `\#12\catcode `\^12\catcode `\_12\catcode `\%12\relax}%
\providecommand \@@startlink[1]{}%
\providecommand \@@endlink[0]{}%
\providecommand \url  [0]{\begingroup\@sanitize@url \@url }%
\providecommand \@url [1]{\endgroup\@href {#1}{\urlprefix }}%
\providecommand \urlprefix  [0]{URL }%
\providecommand \Eprint [0]{\href }%
\providecommand \doibase [0]{https://doi.org/}%
\providecommand \selectlanguage [0]{\@gobble}%
\providecommand \bibinfo  [0]{\@secondoftwo}%
\providecommand \bibfield  [0]{\@secondoftwo}%
\providecommand \translation [1]{[#1]}%
\providecommand \BibitemOpen [0]{}%
\providecommand \bibitemStop [0]{}%
\providecommand \bibitemNoStop [0]{.\EOS\space}%
\providecommand \EOS [0]{\spacefactor3000\relax}%
\providecommand \BibitemShut  [1]{\csname bibitem#1\endcsname}%
\let\auto@bib@innerbib\@empty
\bibitem [{\citenamefont {Colaresi}\ \emph {et~al.}(2021)\citenamefont
  {Colaresi} \emph {et~al.}}]{dresden1}%
  \BibitemOpen
  \bibfield  {author} {\bibinfo {author} {\bibfnamefont {J.}~\bibnamefont
  {Colaresi}} \emph {et~al.},\ }\href@noop {} {\bibfield  {journal} {\bibinfo
  {journal} {Phys. Rev. D}\ }\textbf {\bibinfo {volume} {104}},\ \bibinfo
  {pages} {072003} (\bibinfo {year} {2021})}\BibitemShut {NoStop}%
\bibitem [{\citenamefont {Barbeau}\ \emph {et~al.}(2007)\citenamefont
  {Barbeau}, \citenamefont {Collar},\ and\ \citenamefont {Tench}}]{ppc}%
  \BibitemOpen
  \bibfield  {author} {\bibinfo {author} {\bibfnamefont {P.~S.}\ \bibnamefont
  {Barbeau}}, \bibinfo {author} {\bibfnamefont {J.~I.}\ \bibnamefont
  {Collar}},\ and\ \bibinfo {author} {\bibfnamefont {O.}~\bibnamefont
  {Tench}},\ }\href {https://doi.org/10.1088/1475-7516/2007/09/009} {\bibfield
  {journal} {\bibinfo  {journal} {JCAP}\ }\textbf {\bibinfo {volume}
  {2007}}\bibinfo  {number} { (09)},\ \bibinfo {pages} {009}}\BibitemShut
  {NoStop}%
\bibitem [{\citenamefont {Freedman}(1974)}]{freedman}%
  \BibitemOpen
\bibfield  {number} {  }\bibfield  {author} {\bibinfo {author} {\bibfnamefont
  {D.~Z.}\ \bibnamefont {Freedman}},\ }\href
  {https://doi.org/10.1103/PhysRevD.9.1389} {\bibfield  {journal} {\bibinfo
  {journal} {Phys. Rev. D}\ }\textbf {\bibinfo {volume} {9}},\ \bibinfo {pages}
  {1389} (\bibinfo {year} {1974})}\BibitemShut {NoStop}%
\bibitem [{\citenamefont {Akimov}\ \emph {et~al.}(2017)\citenamefont {Akimov}
  \emph {et~al.}}]{science}%
  \BibitemOpen
  \bibfield  {author} {\bibinfo {author} {\bibfnamefont {D.}~\bibnamefont
  {Akimov}} \emph {et~al.},\ }\href {https://doi.org/10.1126/science.aao0990}
  {\bibfield  {journal} {\bibinfo  {journal} {Science}\ }\textbf {\bibinfo
  {volume} {357}},\ \bibinfo {pages} {1123} (\bibinfo {year}
  {2017})}\BibitemShut {NoStop}%
\bibitem [{\citenamefont {Bowen}\ and\ \citenamefont {Huber}(2020)}]{huber}%
  \BibitemOpen
  \bibfield  {author} {\bibinfo {author} {\bibfnamefont {M.}~\bibnamefont
  {Bowen}}\ and\ \bibinfo {author} {\bibfnamefont {P.}~\bibnamefont {Huber}},\
  }\href {https://doi.org/10.1103/PhysRevD.102.053008} {\bibfield  {journal}
  {\bibinfo  {journal} {Phys. Rev. D}\ }\textbf {\bibinfo {volume} {102}},\
  \bibinfo {pages} {053008} (\bibinfo {year} {2020})}\BibitemShut {NoStop}%
\bibitem [{\citenamefont {Fernandez-Moroni}\ \emph {et~al.}(2021)\citenamefont
  {Fernandez-Moroni} \emph {et~al.}}]{deviations}%
  \BibitemOpen
  \bibfield  {author} {\bibinfo {author} {\bibfnamefont {G.}~\bibnamefont
  {Fernandez-Moroni}} \emph {et~al.},\ }\href@noop {} {} (\bibinfo {year}
  {2021}),\ \Eprint {https://arxiv.org/abs/2108.07310} {arXiv:2108.07310}
  \BibitemShut {NoStop}%
\bibitem [{tri()}]{trip}%
  \BibitemOpen
  \href@noop {} {}\bibinfo {note} {SRCOOL12K Generation 2 portable cooling unit
  by Tripp-Lite, Chicago IL.}\BibitemShut {Stop}%
\bibitem [{\citenamefont {Wong}\ \emph {et~al.}(2007)\citenamefont {Wong} \emph
  {et~al.}}]{wong}%
  \BibitemOpen
  \bibfield  {author} {\bibinfo {author} {\bibfnamefont {H.~T.}\ \bibnamefont
  {Wong}} \emph {et~al.},\ }\href {https://doi.org/10.1103/PhysRevD.75.012001}
  {\bibfield  {journal} {\bibinfo  {journal} {Phys. Rev. D}\ }\textbf {\bibinfo
  {volume} {75}},\ \bibinfo {pages} {012001} (\bibinfo {year}
  {2007})}\BibitemShut {NoStop}%
\bibitem [{\citenamefont {Fernandez~Moroni}\ \emph {et~al.}(2015)\citenamefont
  {Fernandez~Moroni}, \citenamefont {Estrada}, \citenamefont {Paolini},
  \citenamefont {Cancelo}, \citenamefont {Tiffenberg},\ and\ \citenamefont
  {Molina}}]{connie}%
  \BibitemOpen
  \bibfield  {author} {\bibinfo {author} {\bibfnamefont {G.}~\bibnamefont
  {Fernandez~Moroni}}, \bibinfo {author} {\bibfnamefont {J.}~\bibnamefont
  {Estrada}}, \bibinfo {author} {\bibfnamefont {E.~E.}\ \bibnamefont
  {Paolini}}, \bibinfo {author} {\bibfnamefont {G.}~\bibnamefont {Cancelo}},
  \bibinfo {author} {\bibfnamefont {J.}~\bibnamefont {Tiffenberg}},\ and\
  \bibinfo {author} {\bibfnamefont {J.}~\bibnamefont {Molina}},\ }\href
  {https://doi.org/10.1103/PhysRevD.91.072001} {\bibfield  {journal} {\bibinfo
  {journal} {Phys. Rev. D}\ }\textbf {\bibinfo {volume} {91}},\ \bibinfo
  {pages} {072001} (\bibinfo {year} {2015})}\BibitemShut {NoStop}%
\bibitem [{\citenamefont {Beda}\ \emph {et~al.}(2012)\citenamefont {Beda} \emph
  {et~al.}}]{gemma}%
  \BibitemOpen
  \bibfield  {author} {\bibinfo {author} {\bibfnamefont {A.}~\bibnamefont
  {Beda}} \emph {et~al.},\ }\href@noop {} {\bibfield  {journal} {\bibinfo
  {journal} {Adv. High Energy Phys.}\ }\textbf {\bibinfo {volume} {2012}},\
  \bibinfo {pages} {350150} (\bibinfo {year} {2012})}\BibitemShut {NoStop}%
\bibitem [{\citenamefont {Hakenmüller}\ \emph {et~al.}(2019)\citenamefont
  {Hakenmüller} \emph {et~al.}}]{conus2}%
  \BibitemOpen
  \bibfield  {author} {\bibinfo {author} {\bibfnamefont {J.}~\bibnamefont
  {Hakenmüller}} \emph {et~al.},\ }\href
  {https://doi.org/10.1140/epjc/s10052-019-7160-2} {\bibfield  {journal}
  {\bibinfo  {journal} {Eur. Phys. J. C}\ }\textbf {\bibinfo {volume} {79}},\
  \bibinfo {pages} {699} (\bibinfo {year} {2019})}\BibitemShut {NoStop}%
\bibitem [{\citenamefont {Aalseth}\ \emph {et~al.}(2011)\citenamefont {Aalseth}
  \emph {et~al.}}]{surface}%
  \BibitemOpen
  \bibfield  {author} {\bibinfo {author} {\bibfnamefont {C.~E.}\ \bibnamefont
  {Aalseth}} \emph {et~al.},\ }\href
  {https://doi.org/10.1103/PhysRevLett.106.131301} {\bibfield  {journal}
  {\bibinfo  {journal} {Phys. Rev. Lett.}\ }\textbf {\bibinfo {volume} {106}},\
  \bibinfo {pages} {131301} (\bibinfo {year} {2011})}\BibitemShut {NoStop}%
\bibitem [{rel()}]{release}%
  \BibitemOpen
  \href@noop {} {}\bibinfo {note} {Further information can be found in the data
  release accompanying this Letter, which includes Refs.
  \cite{eres1,eres2,fano,toi,wapstra,wilks,Murtaugh,burnham,Akaike,Mundry,Ramsey}.}\BibitemShut
  {Stop}%
\bibitem [{wol()}]{wolfram}%
  \BibitemOpen
  \href@noop {} {}\bibinfo {note} {\textascii{IndependenceTest} built-in
  symbol, Wolfram Language.}\BibitemShut {Stop}%
\bibitem [{\citenamefont {Hakenm\"{u}ller}(2020)}]{thesis}%
  \BibitemOpen
  \bibfield  {author} {\bibinfo {author} {\bibfnamefont {J.~D.}\ \bibnamefont
  {Hakenm\"{u}ller}},\ }\href@noop {} {Ph.D. thesis},\ \bibinfo  {school}
  {Heidelberg University} (\bibinfo {year} {2020})\BibitemShut {NoStop}%
\bibitem [{\citenamefont {Aalseth}\ \emph {et~al.}(2013)\citenamefont {Aalseth}
  \emph {et~al.}}]{cogent}%
  \BibitemOpen
  \bibfield  {author} {\bibinfo {author} {\bibfnamefont {C.~E.}\ \bibnamefont
  {Aalseth}} \emph {et~al.},\ }\href
  {https://doi.org/10.1103/PhysRevD.88.012002} {\bibfield  {journal} {\bibinfo
  {journal} {Phys. Rev. D}\ }\textbf {\bibinfo {volume} {88}},\ \bibinfo
  {pages} {012002} (\bibinfo {year} {2013})}\BibitemShut {NoStop}%
\bibitem [{\citenamefont {Yang}\ \emph {et~al.}(2018)\citenamefont {Yang} \emph
  {et~al.}}]{cdexrt}%
  \BibitemOpen
  \bibfield  {author} {\bibinfo {author} {\bibfnamefont {L.-T.}\ \bibnamefont
  {Yang}} \emph {et~al.},\ }\href@noop {} {\bibfield  {journal} {\bibinfo
  {journal} {Chinese Physics C}\ }\textbf {\bibinfo {volume} {42}},\ \bibinfo
  {pages} {023002} (\bibinfo {year} {2018})}\BibitemShut {NoStop}%
\bibitem [{\citenamefont {Li}\ \emph {et~al.}(2014)\citenamefont {Li} \emph
  {et~al.}}]{cdextanh}%
  \BibitemOpen
  \bibfield  {author} {\bibinfo {author} {\bibfnamefont {H.}~\bibnamefont {Li}}
  \emph {et~al.},\ }\href
  {https://doi.org/https://doi.org/10.1016/j.astropartphys.2014.02.005}
  {\bibfield  {journal} {\bibinfo  {journal} {Astropart. Phys.}\ }\textbf
  {\bibinfo {volume} {56}},\ \bibinfo {pages} {1} (\bibinfo {year}
  {2014})}\BibitemShut {NoStop}%
\bibitem [{\citenamefont {Bonet}\ \emph
  {et~al.}(2021{\natexlab{a}})\citenamefont {Bonet} \emph
  {et~al.}}]{lastconus1}%
  \BibitemOpen
  \bibfield  {author} {\bibinfo {author} {\bibfnamefont {H.}~\bibnamefont
  {Bonet}} \emph {et~al.},\ }\href
  {https://doi.org/10.1103/PhysRevLett.126.041804} {\bibfield  {journal}
  {\bibinfo  {journal} {Phys. Rev. Lett.}\ }\textbf {\bibinfo {volume} {126}},\
  \bibinfo {pages} {041804} (\bibinfo {year} {2021}{\natexlab{a}})}\BibitemShut
  {NoStop}%
\bibitem [{\citenamefont {Bonet}\ \emph
  {et~al.}(2021{\natexlab{b}})\citenamefont {Bonet} \emph
  {et~al.}}]{lastconus2}%
  \BibitemOpen
  \bibfield  {author} {\bibinfo {author} {\bibfnamefont {H.}~\bibnamefont
  {Bonet}} \emph {et~al.},\ }\href
  {https://doi.org/10.1140/epjc/s10052-021-09038-3} {\bibfield  {journal}
  {\bibinfo  {journal} {Eur. Phys. J. C}\ }\textbf {\bibinfo {volume} {81}},\
  \bibinfo {pages} {267} (\bibinfo {year} {2021}{\natexlab{b}})}\BibitemShut
  {NoStop}%
\bibitem [{\citenamefont {Kopeikin}(2012)}]{kopeikin}%
  \BibitemOpen
  \bibfield  {author} {\bibinfo {author} {\bibfnamefont {V.}~\bibnamefont
  {Kopeikin}},\ }\href {https://doi.org/10.1134/S1063778812020123} {\bibfield
  {journal} {\bibinfo  {journal} {Phys. Atom. Nucl.}\ }\textbf {\bibinfo
  {volume} {75}},\ \bibinfo {pages} {143} (\bibinfo {year} {2012})}\BibitemShut
  {NoStop}%
\bibitem [{\citenamefont {Qian}\ and\ \citenamefont {Peng}(2019)}]{review}%
  \BibitemOpen
  \bibfield  {author} {\bibinfo {author} {\bibfnamefont {X.}~\bibnamefont
  {Qian}}\ and\ \bibinfo {author} {\bibfnamefont {J.-C.}\ \bibnamefont
  {Peng}},\ }\href@noop {} {\bibfield  {journal} {\bibinfo  {journal} {Rep.
  Prog. Phys.}\ }\textbf {\bibinfo {volume} {82}},\ \bibinfo {pages} {036201}
  (\bibinfo {year} {2019})}\BibitemShut {NoStop}%
\bibitem [{\citenamefont {Huber}(2011)}]{huber1}%
  \BibitemOpen
  \bibfield  {author} {\bibinfo {author} {\bibfnamefont {P.}~\bibnamefont
  {Huber}},\ }\href@noop {} {\bibfield  {journal} {\bibinfo  {journal} {Phys.
  Rev. C}\ }\textbf {\bibinfo {volume} {84}},\ \bibinfo {pages} {024617}
  (\bibinfo {year} {2011})}\BibitemShut {NoStop}%
\bibitem [{\citenamefont {Mueller}\ \emph {et~al.}(2011)\citenamefont {Mueller}
  \emph {et~al.}}]{mueller}%
  \BibitemOpen
  \bibfield  {author} {\bibinfo {author} {\bibfnamefont {T.~A.}\ \bibnamefont
  {Mueller}} \emph {et~al.},\ }\href@noop {} {\bibfield  {journal} {\bibinfo
  {journal} {Phys. Rev. C}\ }\textbf {\bibinfo {volume} {83}},\ \bibinfo
  {pages} {054615} (\bibinfo {year} {2011})}\BibitemShut {NoStop}%
\bibitem [{\citenamefont {Vogel}\ and\ \citenamefont {Engel}(1989)}]{vogel}%
  \BibitemOpen
  \bibfield  {author} {\bibinfo {author} {\bibfnamefont {P.}~\bibnamefont
  {Vogel}}\ and\ \bibinfo {author} {\bibfnamefont {J.}~\bibnamefont {Engel}},\
  }\href@noop {} {\bibfield  {journal} {\bibinfo  {journal} {Phys. Rev. D}\
  }\textbf {\bibinfo {volume} {39}},\ \bibinfo {pages} {3378} (\bibinfo {year}
  {1989})}\BibitemShut {NoStop}%
\bibitem [{\citenamefont {Liao}\ \emph {et~al.}(2021)\citenamefont {Liao},
  \citenamefont {Liu},\ and\ \citenamefont {Marfatia}}]{danny}%
  \BibitemOpen
  \bibfield  {author} {\bibinfo {author} {\bibfnamefont {J.}~\bibnamefont
  {Liao}}, \bibinfo {author} {\bibfnamefont {H.}~\bibnamefont {Liu}},\ and\
  \bibinfo {author} {\bibfnamefont {D.}~\bibnamefont {Marfatia}},\ }\href@noop
  {} {\bibfield  {journal} {\bibinfo  {journal} {Phys. Rev. D}\ }\textbf
  {\bibinfo {volume} {104}},\ \bibinfo {pages} {015005} (\bibinfo {year}
  {2021})}\BibitemShut {NoStop}%
\bibitem [{dim()}]{dimitris}%
  \BibitemOpen
  \href@noop {} {}\bibinfo {note} {D. Papoulias, private
  communication.}\BibitemShut {Stop}%
\bibitem [{won()}]{wondram}%
  \BibitemOpen
  \href@noop {} {}\bibinfo {note} {2012 Nuclear Data for Reactor Antineutrino
  Measurements Workshop, https://indico.bnl.gov/event/11155/}\BibitemShut
  {NoStop}%
\bibitem [{\citenamefont {Scholz}(2017)}]{bjorn}%
  \BibitemOpen
  \bibfield  {author} {\bibinfo {author} {\bibfnamefont {B.}~\bibnamefont
  {Scholz}},\ }\href@noop {} {Ph.D. thesis},\ \bibinfo  {school} {University of
  Chicago} (\bibinfo {year} {2017}),\ \Eprint
  {https://arxiv.org/abs/1904.01155} {arXiv:1904.01155} \BibitemShut {NoStop}%
\bibitem [{\citenamefont {Lindhard}\ \emph {et~al.}(1963)\citenamefont
  {Lindhard}, \citenamefont {Nielsen}, \citenamefont {Scharff},\ and\
  \citenamefont {Thomsen}}]{lindhard}%
  \BibitemOpen
  \bibfield  {author} {\bibinfo {author} {\bibfnamefont {J.}~\bibnamefont
  {Lindhard}}, \bibinfo {author} {\bibfnamefont {V.}~\bibnamefont {Nielsen}},
  \bibinfo {author} {\bibfnamefont {M.}~\bibnamefont {Scharff}},\ and\ \bibinfo
  {author} {\bibfnamefont {P.~V.}\ \bibnamefont {Thomsen}},\ }\href
  {https://www.osti.gov/biblio/4701226} {\bibfield  {journal} {\bibinfo
  {journal} {Kgl. Danske Videnskab., Selskab. Mat. Fys. Medd.}\ }\textbf
  {\bibinfo {volume} {33}},\ \bibinfo {pages} {10} (\bibinfo {year}
  {1963})}\BibitemShut {NoStop}%
\bibitem [{\citenamefont {Collar}\ \emph {et~al.}(2021)\citenamefont {Collar},
  \citenamefont {Kavner},\ and\ \citenamefont {Lewis}}]{qf}%
  \BibitemOpen
  \bibfield  {author} {\bibinfo {author} {\bibfnamefont {J.~I.}\ \bibnamefont
  {Collar}}, \bibinfo {author} {\bibfnamefont {A.~R.~L.}\ \bibnamefont
  {Kavner}},\ and\ \bibinfo {author} {\bibfnamefont {C.~M.}\ \bibnamefont
  {Lewis}},\ }\href {https://doi.org/10.1103/PhysRevD.103.122003} {\bibfield
  {journal} {\bibinfo  {journal} {Phys. Rev. D}\ }\textbf {\bibinfo {volume}
  {103}},\ \bibinfo {pages} {122003} (\bibinfo {year} {2021})}\BibitemShut
  {NoStop}%
\bibitem [{\citenamefont {Agnese}\ \emph {et~al.}(2016)\citenamefont {Agnese}
  \emph {et~al.}}]{supercdms}%
  \BibitemOpen
  \bibfield  {author} {\bibinfo {author} {\bibfnamefont {R.}~\bibnamefont
  {Agnese}} \emph {et~al.},\ }\href
  {https://doi.org/10.1103/PhysRevLett.116.071301} {\bibfield  {journal}
  {\bibinfo  {journal} {Phys. Rev. Lett.}\ }\textbf {\bibinfo {volume} {116}},\
  \bibinfo {pages} {071301} (\bibinfo {year} {2016})}\BibitemShut {NoStop}%
\bibitem [{\citenamefont {Schönfeld}(1998)}]{theory}%
  \BibitemOpen
  \bibfield  {author} {\bibinfo {author} {\bibfnamefont {E.}~\bibnamefont
  {Schönfeld}},\ }\href@noop {} {\bibfield  {journal} {\bibinfo  {journal}
  {Appl. Radiat. Isot.}\ }\textbf {\bibinfo {volume} {49}},\ \bibinfo {pages}
  {1353} (\bibinfo {year} {1998})}\BibitemShut {NoStop}%
\bibitem [{\citenamefont {Mougeot}(2019)}]{mougeot1}%
  \BibitemOpen
  \bibfield  {author} {\bibinfo {author} {\bibfnamefont {X.}~\bibnamefont
  {Mougeot}},\ }\href@noop {} {\bibfield  {journal} {\bibinfo  {journal} {Appl.
  Radiat. Isot.}\ }\textbf {\bibinfo {volume} {154}},\ \bibinfo {pages}
  {108884} (\bibinfo {year} {2019})}\BibitemShut {NoStop}%
\bibitem [{mou()}]{mougeot2}%
  \BibitemOpen
  \href@noop {} {}\bibinfo {note} {X. Mougeot, private
  communication.}\BibitemShut {Stop}%
\bibitem [{\citenamefont {Foreman-Mackey}\ \emph {et~al.}(2013)\citenamefont
  {Foreman-Mackey}, \citenamefont {Hogg}, \citenamefont {Lang},\ and\
  \citenamefont {Goodman}}]{mcmc1}%
  \BibitemOpen
  \bibfield  {author} {\bibinfo {author} {\bibfnamefont {D.}~\bibnamefont
  {Foreman-Mackey}}, \bibinfo {author} {\bibfnamefont {D.~W.}\ \bibnamefont
  {Hogg}}, \bibinfo {author} {\bibfnamefont {D.}~\bibnamefont {Lang}},\ and\
  \bibinfo {author} {\bibfnamefont {J.}~\bibnamefont {Goodman}},\ }\href
  {https://doi.org/10.1086/670067} {\bibfield  {journal} {\bibinfo  {journal}
  {Publ. Astron. Soc. Pac.}\ }\textbf {\bibinfo {volume} {125}},\ \bibinfo
  {pages} {306} (\bibinfo {year} {2013})}\BibitemShut {NoStop}%
\bibitem [{mcm()}]{mcmc2}%
  \BibitemOpen
  \href@noop {} {}\bibinfo {howpublished}
  {\url{https://emcee.readthedocs.io/en/v2.2.1/}}\BibitemShut {NoStop}%
\bibitem [{\citenamefont {Kass}\ and\ \citenamefont {Raftery}(1995)}]{bayes}%
  \BibitemOpen
  \bibfield  {author} {\bibinfo {author} {\bibfnamefont {R.~E.}\ \bibnamefont
  {Kass}}\ and\ \bibinfo {author} {\bibfnamefont {A.~E.}\ \bibnamefont
  {Raftery}},\ }\href {http://www.jstor.org/stable/2291091} {\bibfield
  {journal} {\bibinfo  {journal} {J. Am. Stat. Assoc.}\ }\textbf {\bibinfo
  {volume} {90}},\ \bibinfo {pages} {773} (\bibinfo {year} {1995})}\BibitemShut
  {NoStop}%
\bibitem [{\citenamefont {Jeffreys}(1961)}]{jeffreys}%
  \BibitemOpen
  \bibfield  {author} {\bibinfo {author} {\bibfnamefont {H.}~\bibnamefont
  {Jeffreys}},\ }\href@noop {} {\emph {\bibinfo {title} {Theory of
  Probability}}},\ \bibinfo {edition} {3rd}\ ed.\ (\bibinfo  {publisher}
  {Oxford},\ \bibinfo {address} {Oxford, England},\ \bibinfo {year}
  {1961})\BibitemShut {NoStop}%
\bibitem [{\citenamefont {Collar}\ \emph {et~al.}(2019)\citenamefont {Collar},
  \citenamefont {Kavner},\ and\ \citenamefont {Lewis}}]{csiqf}%
  \BibitemOpen
  \bibfield  {author} {\bibinfo {author} {\bibfnamefont {J.~I.}\ \bibnamefont
  {Collar}}, \bibinfo {author} {\bibfnamefont {A.~R.~L.}\ \bibnamefont
  {Kavner}},\ and\ \bibinfo {author} {\bibfnamefont {C.~M.}\ \bibnamefont
  {Lewis}},\ }\href {https://doi.org/10.1103/PhysRevD.100.033003} {\bibfield
  {journal} {\bibinfo  {journal} {Phys. Rev. D}\ }\textbf {\bibinfo {volume}
  {100}},\ \bibinfo {pages} {033003} (\bibinfo {year} {2019})}\BibitemShut
  {NoStop}%
\bibitem [{\citenamefont {Bonhomme}\ \emph {et~al.}()\citenamefont {Bonhomme}
  \emph {et~al.}}]{conusqf}%
  \BibitemOpen
  \bibfield  {author} {\bibinfo {author} {\bibfnamefont {A.}~\bibnamefont
  {Bonhomme}} \emph {et~al.},\ }\href@noop {} {}\Eprint
  {https://arxiv.org/abs/2202.03754} {arXiv:2202.03754} \BibitemShut {NoStop}%
\bibitem [{\citenamefont {Collar}\ and\ \citenamefont {Lewis}()}]{ourcom}%
  \BibitemOpen
  \bibfield  {author} {\bibinfo {author} {\bibfnamefont {J.~I.}\ \bibnamefont
  {Collar}}\ and\ \bibinfo {author} {\bibfnamefont {C.~M.}\ \bibnamefont
  {Lewis}},\ }\href@noop {} {}\Eprint {https://arxiv.org/abs/2203.00750}
  {arXiv:2203.00750} \BibitemShut {NoStop}%
\bibitem [{\citenamefont {Barbeau}(2009)}]{phil}%
  \BibitemOpen
  \bibfield  {author} {\bibinfo {author} {\bibfnamefont {P.}~\bibnamefont
  {Barbeau}},\ }\href@noop {} {Ph.D. thesis},\ \bibinfo  {school} {University
  of Chicago} (\bibinfo {year} {2009})\BibitemShut {NoStop}%
\bibitem [{\citenamefont {Baxter}\ \emph {et~al.}(2020)\citenamefont {Baxter}
  \emph {et~al.}}]{ESS}%
  \BibitemOpen
  \bibfield  {author} {\bibinfo {author} {\bibfnamefont {D.}~\bibnamefont
  {Baxter}} \emph {et~al.},\ }\href {https://doi.org/10.1007/JHEP02(2020)123}
  {\bibfield  {journal} {\bibinfo  {journal} {J. High Energy Phys}\ }\textbf
  {\bibinfo {volume} {2020}},\ \bibinfo {pages} {123} (\bibinfo {year}
  {2020})}\BibitemShut {NoStop}%
\bibitem [{\citenamefont {Aalseth}\ \emph {et~al.}(2008)\citenamefont {Aalseth}
  \emph {et~al.}}]{eres1}%
  \BibitemOpen
  \bibfield  {author} {\bibinfo {author} {\bibfnamefont {C.~E.}\ \bibnamefont
  {Aalseth}} \emph {et~al.},\ }\href@noop {} {\bibfield  {journal} {\bibinfo
  {journal} {Phys. Rev. Lett.}\ }\textbf {\bibinfo {volume} {101}},\ \bibinfo
  {pages} {251301} (\bibinfo {year} {2008})}\BibitemShut {NoStop}%
\bibitem [{\citenamefont {Aalseth}\ \emph {et~al.}(2009)\citenamefont {Aalseth}
  \emph {et~al.}}]{eres2}%
  \BibitemOpen
  \bibfield  {author} {\bibinfo {author} {\bibfnamefont {C.~E.}\ \bibnamefont
  {Aalseth}} \emph {et~al.},\ }\href@noop {} {\bibfield  {journal} {\bibinfo
  {journal} {Phys. Rev. Lett.}\ }\textbf {\bibinfo {volume} {102}},\ \bibinfo
  {pages} {109903} (\bibinfo {year} {2009})}\BibitemShut {NoStop}%
\bibitem [{\citenamefont {Lowe}(1997)}]{fano}%
  \BibitemOpen
  \bibfield  {author} {\bibinfo {author} {\bibfnamefont {B.}~\bibnamefont
  {Lowe}},\ }\href@noop {} {\bibfield  {journal} {\bibinfo  {journal} {Nucl.
  Instr. Meth. A}\ }\textbf {\bibinfo {volume} {399}},\ \bibinfo {pages} {354}
  (\bibinfo {year} {1997})}\BibitemShut {NoStop}%
\bibitem [{\citenamefont {Firestone}\ and\ \citenamefont
  {Shirley}(1996)}]{toi}%
  \BibitemOpen
  \bibfield  {author} {\bibinfo {author} {\bibfnamefont {R.~B.}\ \bibnamefont
  {Firestone}}\ and\ \bibinfo {author} {\bibfnamefont {V.~S.}\ \bibnamefont
  {Shirley}},\ }\href@noop {} {\emph {\bibinfo {title} {Table of Isotopes}}}\
  (\bibinfo  {publisher} {Wiley},\ \bibinfo {address} {New York},\ \bibinfo
  {year} {1996})\BibitemShut {NoStop}%
\bibitem [{\citenamefont {Konopinski}\ and\ \citenamefont
  {Rose}(1965)}]{wapstra}%
  \BibitemOpen
  \bibfield  {author} {\bibinfo {author} {\bibfnamefont {E.}~\bibnamefont
  {Konopinski}}\ and\ \bibinfo {author} {\bibfnamefont {M.}~\bibnamefont
  {Rose}},\ }\bibfield  {title} {\bibinfo {title} {The theory of nuclear
  $\beta$-decay},\ }in\ \href@noop {} {\emph {\bibinfo {booktitle} {Alpha-,
  Beta- and Gamma-Ray Spectroscopy}}},\ \bibinfo {editor} {edited by\ \bibinfo
  {editor} {\bibfnamefont {K.}~\bibnamefont {Siegbahn}}}\ (\bibinfo
  {publisher} {North-Holland},\ \bibinfo {address} {Amsterdam},\ \bibinfo
  {year} {1965})\BibitemShut {NoStop}%
\bibitem [{\citenamefont {Wilks}(1938)}]{wilks}%
  \BibitemOpen
  \bibfield  {author} {\bibinfo {author} {\bibfnamefont {S.~S.}\ \bibnamefont
  {Wilks}},\ }\href@noop {} {\bibfield  {journal} {\bibinfo  {journal} {Ann.
  Math. Stat.}\ }\textbf {\bibinfo {volume} {9}},\ \bibinfo {pages} {60 }
  (\bibinfo {year} {1938})}\BibitemShut {NoStop}%
\bibitem [{\citenamefont {Murtaugh}(2014)}]{Murtaugh}%
  \BibitemOpen
  \bibfield  {author} {\bibinfo {author} {\bibfnamefont {P.~A.}\ \bibnamefont
  {Murtaugh}},\ }\href@noop {} {\bibfield  {journal} {\bibinfo  {journal}
  {Ecology}\ }\textbf {\bibinfo {volume} {95}},\ \bibinfo {pages} {611}
  (\bibinfo {year} {2014})}\BibitemShut {NoStop}%
\bibitem [{\citenamefont {Burnham}\ and\ \citenamefont
  {Anderson}(2003)}]{burnham}%
  \BibitemOpen
  \bibfield  {author} {\bibinfo {author} {\bibfnamefont {K.}~\bibnamefont
  {Burnham}}\ and\ \bibinfo {author} {\bibfnamefont {D.}~\bibnamefont
  {Anderson}},\ }\href@noop {} {\emph {\bibinfo {title} {Model Selection and
  Multimodel Inference: A Practical Information-Theoretic Approach}}}\
  (\bibinfo  {publisher} {Springer New York},\ \bibinfo {year}
  {2003})\BibitemShut {NoStop}%
\bibitem [{\citenamefont {{Akaike}}(1974)}]{Akaike}%
  \BibitemOpen
  \bibfield  {author} {\bibinfo {author} {\bibfnamefont {H.}~\bibnamefont
  {{Akaike}}},\ }\href@noop {} {\bibfield  {journal} {\bibinfo  {journal} {IEEE
  Trans. Automat. Contr.}\ }\textbf {\bibinfo {volume} {19}},\ \bibinfo {pages}
  {716} (\bibinfo {year} {1974})}\BibitemShut {NoStop}%
\bibitem [{\citenamefont {Mundry}(2010)}]{Mundry}%
  \BibitemOpen
  \bibfield  {author} {\bibinfo {author} {\bibfnamefont {R.}~\bibnamefont
  {Mundry}},\ }\href@noop {} {\bibfield  {journal} {\bibinfo  {journal} {Behav.
  Ecol. Sociobiol.}\ }\textbf {\bibinfo {volume} {65}},\ \bibinfo {pages} {57}
  (\bibinfo {year} {2010})}\BibitemShut {NoStop}%
\bibitem [{\citenamefont {Ramsey}\ and\ \citenamefont
  {Schafer}(2002)}]{Ramsey}%
  \BibitemOpen
  \bibfield  {author} {\bibinfo {author} {\bibfnamefont {F.~L.}\ \bibnamefont
  {Ramsey}}\ and\ \bibinfo {author} {\bibfnamefont {D.~W.}\ \bibnamefont
  {Schafer}},\ }\href@noop {} {\emph {\bibinfo {title} {The Statistical Sleuth:
  A Course in Methods of Data Analysis}}}\ (\bibinfo  {publisher} {Brooks/Cole,
  Cengage Learning},\ \bibinfo {year} {2002})\BibitemShut {NoStop}%
\end{thebibliography}%

\end{document}